\documentclass[9pt,twocolumn,twoside]{pnas-new}

\templatetype{pnasresearcharticle} 

\usepackage[utf8]{inputenc} 
\usepackage[T1]{fontenc}    
\usepackage{hyperref}       
\usepackage{url}            
\usepackage{booktabs}       
\usepackage{amsfonts}       
\usepackage{nicefrac}       
\usepackage{graphicx}

\graphicspath{ {figures/} }

\hypersetup{
pdftitle={igraph enables fast and robust network analysis across programming languages},
pdfsubject={q-bio.NC, q-bio.QM},
pdfauthor={Michael Antonov, Szabolcs Horvát, Kirill Müller, Tamás Nepusz, Daniel Noom, Vincent Traag, Brooke Foucault Welles, Fabio Zanini},
pdfkeywords={network analysis, open source, graphs, mathematics},
}

\begin{document}

\title{igraph enables fast and robust network analysis across programming languages}


\author[1,*,+]{Michael Antonov}
\author[2,*]{Gábor Csárdi}
\author[3,4,*,+]{Szabolcs Horvát}
\author[1,*,+]{Kirill Müller}
\author[5,*,+]{Tamás Nepusz}
\author[6,*,+]{Daniel Noom}
\author[1,*]{Maëlle Salmon}
\author[7,*,+]{Vincent Traag}
\author[8,*,+]{Brooke Foucault Welles}
\author[9,10,*,+]{Fabio Zanini}

\affil[1]{cynkra GmbH, Zürich, Switzerland}
\affil[2]{Posit Software, PBC, Boston, MA, 02210, United States of America}
\affil[3]{Reykjavik University, Reykjavík, Iceland}
\affil[4]{Center for Systems Biology Dresden, Dresden, Germany}
\affil[5]{Sixdegrees Ltd, Budapest, Hungary}
\affil[6]{jitjit, Amsterdam, Netherlands}
\affil[7]{Centre for Science and Technology Studies, Leiden University, Leiden, Zuid-Holand, Netherlands}
\affil[8]{Northeastern University, Boston, MA, 02115, United States of America}
\affil[9]{School of Clinical Medicine, UNSW Sydney, NSW, 2052, Australia}
\affil[10]{Evolution \& Ecology Research Centre, UNSW Sydney, NSW, 2052, Australia}


\authorcontributions{C core: GC, SzH, TN, DN, VT, FZ. Python: TN, VT, FZ. R: GC, MA, SzH, KM, TN, MS, VT. Mathematica: SzH. Community: VT, BFW.}
\authordeclaration{The authors do not declare any competing interests.}
\equalauthors{\textsuperscript{*} All authors contributed equally to this work.}
\correspondingauthor{

\textsuperscript{+}To whom correspondence should be addressed. E-mail:  szhorvat@gmail.com, antonov5551998@gmail.com, kirill@cynkra.com, ntamas@protonmail.ch, dannoom@hotmail.com, v.a.traag@cwts.leidenuniv.nl, b.welles@northeastern.edu, fabio.zanini@unsw.edu.au
}

\begin{abstract}
Networks or graphs are widely used across the sciences to represent relationships of many kinds. \texttt{igraph} (\url{https://igraph.org}) is a general-purpose software library for graph construction, analysis, and visualisation, combining fast and robust performance with a low entry barrier. \texttt{igraph} pairs a fast core written in C with beginner-friendly interfaces in Python, R, and Mathematica. Over the last two decades, \texttt{igraph} has expanded substantially. It now scales to billions of edges, supports Mathematica and interactive plotting, integrates with Jupyter notebooks and other network libraries, includes new graph layouts and community detection algorithms, and has streamlined the documentation with examples and Spanish translations. Modern testing features such as continuous integration, address sanitizers, stricter typing, and memory-managed vectors have also increased robustness. Hundreds of bug reports have been fixed and a community forum has been opened to connect users and developers. Specific effort has been made to broaden use and community participation by women, non-binary people, and other demographic groups typically underrepresented in open source software.  
\end{abstract}


\dates{This manuscript was compiled on \today}

\maketitle
\thispagestyle{firststyle}
\ifthenelse{\boolean{shortarticle}}{\ifthenelse{\boolean{singlecolumn}}{\abscontentformatted}{\abscontent}}{}

\significancestatement{
Graphs or networks are used to represent relationships of all kinds. Here, we present 17 years of development of \texttt{igraph}, a popular software library for computational analysis of networks in four programming languages for a vast audience in mathematics, computer science, engineering, biology, medicine, and social sciences. The library provides hundreds of functions for network construction, structural analysis, community detection, 2D and 3D visualisation, and storage. \texttt{igraph} is used by hundreds of packages and downloaded 3 million times each month. Development and governance follow transparent, open science principles and encourage community engagement, especially by members of underrepresented groups in network science. The code is entirely open source and  all contributions are welcome.
}

\firstpage{2}
\dropcap{N}etworks or graphs are widely used concepts across multiple scientific disciplines. They allow researcher to study not only the constituent elements that are of interest to their disciplines, but also the interaction between those elements. In mathematics, graphs have been studied for centuries from a theoretical perspective and are still a topic of considerable interest. Random graphs, initially defined by Erd\H{o}s and R\'enyi \citep{erdos59a}, and more recently scale-free stochastic networks, have sparked a new wave of interest aimed at understanding the analytical properties of graphs generated according to statistical rules, with real world applications in engineering, finance, and biology \citep{Barabasi1999-uz}. In biology and medicine, the genomics revolution driven by next-generation sequencing has brought graphs into the spotlight for a wide array of applications, from De Bruijn graphs for genome assembly \citep{Bankevich2012-um} to cell similarity networks in cell atlases \citep{Tabula_Muris_Consortium2018-yg}. Social network analysis has a long history, dating back to Moreno's sociometry \citep{Moreno1941-kk}, and includes the widespread study of social systems, ranging from friendships to organizations and social movements \citep{Burt2000-ne,Diani1992-bz,McFarland2014-iz}. The advent of large online behavioural data opened up many new possibilities, helping give rise to computational social science (Lazer et al.), in which networks play a pivotal role.

Open source computational tools are of central importance in the investigation of graphs. In the case of random graphs, computers enable the simulation of thousands of such graphs at scale and allow empirical computations of their statistical properties. Real-world networks such as protein-protein interactions, single-cell transcriptomics, citation networks or online social networks have thousands or even millions of vertices. High-performance and efficient software libraries are therefore absolutely necessary for their analysis and manipulation. A few general-purpose graph analysis libraries are widely used. NetworkX is a popular library in the Python community \citep{networkx} that is written in pure Python. Graph-tool is also used by many researchers for its faster speed compared to NetworkX and its focus on stochastic networks \citep{peixoto_graph-tool_2014}. StatNet is an R library geared towards dynamic networks and statistical inference that is popular among social scientists \citep{statnet2003}. In addition to graph analysis, graph visualisation is also essential across multiple scientific disciplines. Cytoscape is a popular choice for advanced graph visualization \citep{Shannon2003-qa}, while D3.js can be used from JavaScript to embed networks in a web page \citep{d3js}. Gephi is a popular network analysis tool with a central role for graph visualization \citep{gephi}, and both NetworkX and graph-tool can produce relatively simple visualisations from Python. Despite this vibrant software ecosystem, there is a strong demand for a general-purpose graph analysis library that is fast, consistent across programming languages, cross-platform, easy to install, and well documented. \texttt{igraph} meets these needs by providing efficient implementations of widely used algorithms, many of which are not available in any other library, including for determining shortest paths \citep{Yen1970-pn}, vertex centrality \citep{Brandes2011-rw,Gleich_2005-me}, cycle bases \citep{horton}, motif finding \citep{Wernicke2006-yz}, clique finding \citep{Ostergard2002-eg}, community detection \citep{Traag2019-xy}, and graph layout \citep{McInnes2018-eu}.

\section*{Results}
\label{sec:results}

\subsection*{Growing a large userbase across four programming languages}
Since its first publication in 2006 \citep{igraph1}, \texttt{igraph} has grown in multiple areas, including the C core, extensions in Python, R, and Mathematica \citep{Horvat2023-wh}. Overall, the codebase has had around 3 million lines of code edited spread across approximately 10,000 versioning commits. The four main code repositories, one for each programming language, accumulated a total of around 800 forks and 3,000 stars. The Python and R packages were downloaded around 1 million and 600 thousand times respectively in the last month, which is a conservative estimate given that tracking all downloads is virtually impossible.

\subsection*{Scaling network analysis to billion of edges}
\texttt{igraph} is based on a low-level core codebase in C, on top of which higher-level language interfaces are built. The C core has grown to approximately 1,100 user-facing functions and has become substantially more robust than in previous versions. The build system has been overhauled and now works with CMake, allowing easy cross-platform installation from source in Linux, macOS and Windows. Precompiled packages are also available at the system level (e.g. Ubuntu, Debian, and Alpine Linux, Homebrew and MacPorts for macOS, and \texttt{vcpkg} for Windows) and the language-specific extensions are available on PyPI for Python, the comprehensive R network (CRAN) for R, and as a Paclet for Mathematica.

\begin{figure}[ht]
\includegraphics[width=8cm]{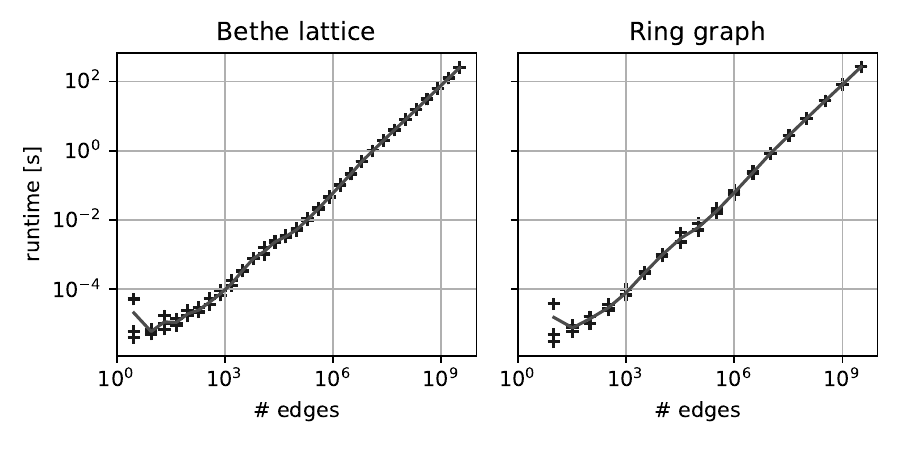}
\centering
\caption{Runtimes to construct an undirected Bethe lattice or regular tree (left) and a directed ring graph (right) using the new \texttt{igraph} C core on a workstation (single CPU). Each graph size was tested three times (crosses). Average of each triplet are shown as solid lines.}
\label{fig:scaling}
\end{figure}

The package now supports 64-bit integers throughout, enabling analysis of networks with billions of vertices and edges. As an example, constructing an undirected Bethe lattice with 3.2 billion edges took 4 minutes on a tower workstation using a single CPU core (Figure~\ref{fig:scaling}, left). Similar results were obtained when building a directed ring graph with 3.3 billion edges (Figure~\ref{fig:scaling}, right). The Python and Mathematica interfaces also support 64-bit integers. The foundations for backward-compatible 64-bit support in R have been laid, using doubles instead of 64-bit integers due to R's current internal limitations. The C core also offers memory managed vector-like data structures, which simplifies internal code and enables users to more easily manage arrays of data, such as vectors of vertices and edges.

New graph algorithms have also been added to the C core and interfaces. Community detection, one the most common tasks of network analysis, now includes additional algorithms such as Leiden \citep{Traag2019-xy}, multilevel (or Louvain) \citep{Blondel2008-ht} and fluid communities \citep{fluidcommunities}. A number of graph layout algorithms have been added, including Uniform Manifold Approximation and Projection (UMAP) \citep{McInnes2018-eu}, or improved, including Kamada-Kawai. Importantly, this implementation of UMAP does not depend on cross-language packages such as \texttt{numba} and \texttt{llvmlite}, which can sometimes cause installation issues. Other layout algorithms available in \texttt{igraph} include Fruchterman-Reingold, multidimensional scaling \citep{Cox2008-lr}, the Large Graph Layout \citep{Adai2004-ba}, Davidson-Harel \citep{Davidson1996-vz}, and the GEM layout \citep{Frick1995-eg}.

\subsection*{Broad user accessibility in R, Python, and Mathematica}
Many aspects of \texttt{igraph}'s high-level interfaces have also been improved.

\begin{figure}[ht]
\includegraphics[width=8.5cm]{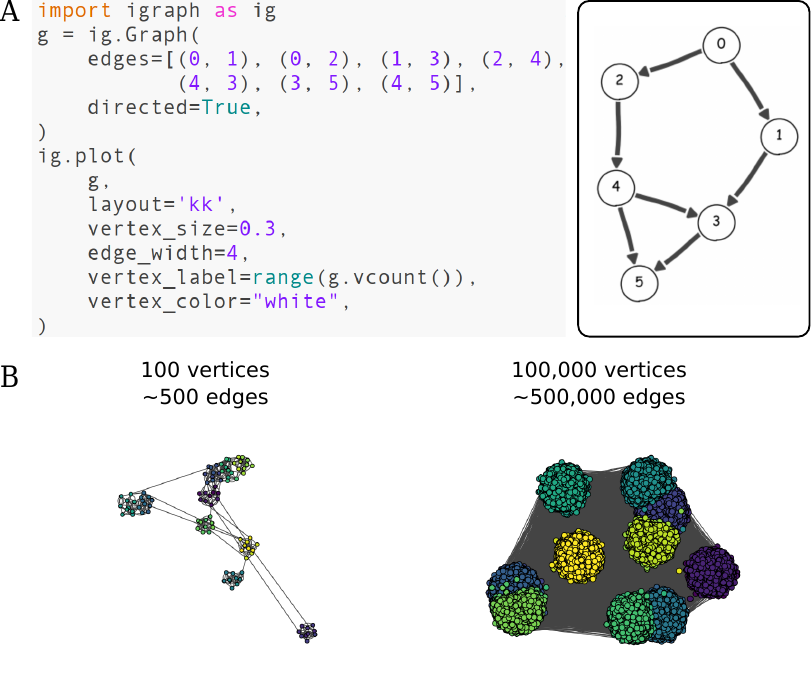}
\centering
\caption{\texttt{igraph} enables simple and interactive graph visualisation. (\textbf{A}) Example code to plot a simple graph with topological sorting. (\textbf{B}) \texttt{igraph} implementation of the Universal Manifold Approximation and Projection (UMAP) layout algorithm \citep{McInnes2018-eu} for a set of 10 groups of vertices that are tightly connected internally and with random distance between groups (See Supplementary Information - listing 1). All plots use the new \texttt{matplotlib} backend for \texttt{igraph}.}
\label{fig:1}
\end{figure}

\begin{figure}[ht]
\includegraphics[width=8.5cm]{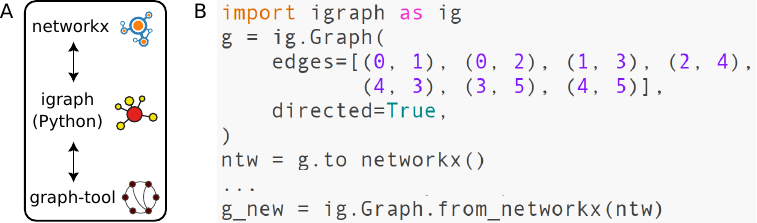}
\centering
\caption{Linking network libraries together. \textbf{(A)} \texttt{igraph}'s Python interface can exchange in-memory graphs with networkx \citep{networkx} and graph-tool \citep{peixoto_graph-tool_2014}. \textbf{(B)} Example of a network conversion from \texttt{igraph} to \texttt{networkx} and back using the new import/export functions, which preserve attributes as well.}
\label{fig:2}
\end{figure}

The Python interface, which includes around 600 functions, is now compatibile with Jupyter notebooks, a browser-based environment for scientific computing and rapid prototyping that supports both R and Python and is widely used in education, research, and industry \citep{Granger2021-rz}. The Python visualisation layer has also been redesigned to support not only the existing \texttt{Cairo} backend but also \texttt{matplotlib} \citep{Hunter:2007} and \texttt{plotly} \citep{plotly}, which enable interactive plots (e.g. zoom and pan) and animations (Figure~\ref{fig:1}). Furthermore, \texttt{igraph}'s Python interface supports in-memory import/export of graphs to and from \texttt{networkx} \citep{networkx} and \texttt{graph-tool} \citep{peixoto_graph-tool_2014}, two popular network analysis libraries (Figure~\ref{fig:2}A). Conversions preserve graph, vertex, and edge attributes, making it easy to convert an \texttt{igraph} graph into a \texttt{networkx} object, call a specific function on it, then convert back to \texttt{igraph} for further analysis (Figure~\ref{fig:2}B). Compatibility with pandas dataframes has been added as well \citep{mckinney-proc-scipy-2010}.

The R interface of \texttt{igraph} is now compatible with R versions 4.0 and later and has expanded to around 750 user functions. On CRAN, the main code repository for the R language, \texttt{igraph} is imported by 713 packages: \texttt{igraph} has even been used just this year to compute CRAN's own dependency graph \citep{crandep}. Many new C functions have been exposed to R including new community detection methods and layout algorithms. Moreover, both the user interface and the structure of the package are now more idiomatic to R and the package ecosystem, lowering the entry barrier for contributors. A number of bug fixes and inconsistencies that accumulated over the years have also been solved.

A Mathematica interface has been added, available at \url{http://szhorvat.net/mathematica/IGraphM}. Unlike the C, R, and Python interfaces, it is mainly designed to extend the network analysis capabilities of the Mathematica standard library rather than to replace them. For instance, the Mathematica interface exposes useful community detection, layout, motif finding, and random graph generation algorithms that are otherwise not available. It also implements a dedicated interactive editor to construct graphs visually, which is especially appealing for prototyping or brainstorming. Details of the Mathematica interface were described in a recent publication \citep{Horvat2023-wh}.

\begin{figure}[ht]
\includegraphics[width=7.5cm]{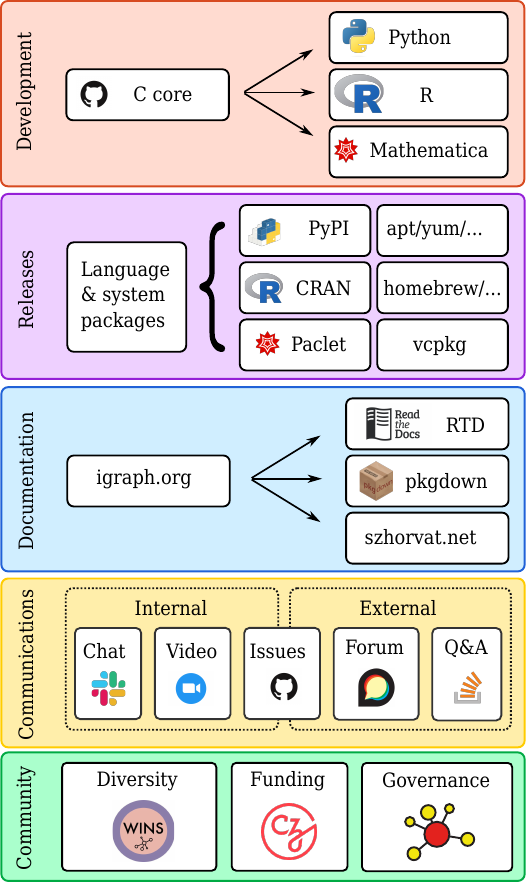}
\centering
\caption{Schematic of \texttt{igraph}'s open source/open science infrastructure. From the top: development, releases, documentation, internal and external communication channels, and initiatives.}
\label{fig:3}
\end{figure}

\subsection*{Refreshed documentation and testing}
In addition to source code changes, \texttt{igraph} has now a more robust testing framework which has led to addressing hundreds of bug reports and enhancement requests. Continuous integration across multiple architectures (64-bit Linux for Intel and AMD CPUs, macOS for 64-bit Intel and the new M1/M2 CPUs, and Windows), including usage of address sanitizers, ensures that the codebase is not just well tested but that new contributions are thoroughly vetted as well. Test coverage for the C core is 85\% excluding vendored dependencies which are already tested separately. In the Python interface, visualization functions are now tested by pixel-by-pixel matching of the resulting images, as in \texttt{matplotlib} \citep{Hunter:2007}.

\texttt{igraph}'s documentation has also improved significantly. In R, a new package vignette and documentation website using \texttt{pkgdown} \citep{pkgdown} were designed to lower the entry barrier for new users. The Python interface is now provided through \texttt{readthedocs.org} and includes a gallery of examples available as scripts or Jupyter notebooks \citep{oscar-najera}. Translations for both the R and Python interface tutorials into Spanish have been added, and other translations (e.g. Mandarin) are planned.

\subsection*{Open science development towards a diverse community}
We have also aimed to increase contact with the user community and participation into the development itself (Figure~\ref{fig:3}). The team has committed to a transparent development process. GitHub issues are the agora where aspiring contributors interact with the core development team on specific bugs, feature requests, or enhancement proposals, which are then turned into public pull requests. Automatic testing via continuous integration is configured openly, with results accessible by anyone in the world. Releases are triggered for the popular dissemination platforms for each language (e.g. PyPI for Python, CRAN for R) and relationships with downstream package managers are actively maintained. The \texttt{igraph} documentation is now deployed using language-specific conventions, which feels familiar even to new users. In addition to internal communication on development via chat rooms and regular video calls, \texttt{igraph} now has a public user forum at \url{https://igraph.discourse.group/} that is regularly visited by both developers and users. The developers also check in on online Q\&A platforms such as Stack Overflow to provide answers.

The recent collaboration with Women in Network Science (WiNS) is aiming to extend our contributor base to a more diverse group of individuals, especially from women, non-binary people and people from other groups that are traditionally underrepresented in open source development teams. A team led by Foucault Welles has completed a study of the igraph forums and suggested best practices for on-boarding and retaining new users. The team is currently conducting interviews with open source community managers on strategies for increasing diversity, inclusion and belonging in open source communities. In addition, a recent workshop on network analysis using \texttt{igraph} took place at Northeastern University in Boston and was inspirational as a blueprint for further community engagement in the future. While there is much work ahead, we remain deeply committed to increasing diversity among both the user and developer community behind \texttt{igraph} to ensure our code, documentation, and governance are representing all humans independent of gender, race, or background. Finally, \texttt{igraph} is in the process of setting up a transparent governance structure, aimed at both clarifying and simplifying the decision making process and improve the relationship between core developers and the user community.

\section*{Discussion}
\label{sec:discussion}
The computational analysis of graphs and networks has become a foundational element across scientific disciplines including mathematics, physics, engineering, biology and medicine, and the social sciences. It is also essential in industry settings, such as graph-based machine learning. A general-purpose, open source network analysis software library is essential for academia, start-up companies, non-profit organisations, and for other small teams and individuals. Over the last decade, \texttt{igraph} has morphed from a remarkable hobby project of two developers into an essential tool across disciplines and industries.

\texttt{igraph}'s scalability to billions of edges on consumer hardware is poised to simplify analytics on important graphs that are challenging due to their sheer size. Huge networks, which include a Twitter network of 40 million vertices and 1.5 billion edges \citep{Kwak2010-zh} and a citation network with more than 240 million vertices and almost 2 billion edges \citep{sinha2015an}, have become increasingly common in recent years. Beyond current data capabilities, \texttt{igraph} is ready for future analyses of global networks, for instance in epidemiology (8 billion people) and on social media (e.g. Facebook, 3 billion users). In turn, the results of such massive computations can inform key decisions in human society, for instance to contain viral outbreaks and combat hate speech.

Interfaces in high-level languages have seen widespread use across the sciences due to their relatively flat learning curve and affinity for exploratory data science. In particular, systems biomedicine is a growing field where networks are of paramount importance. Cell-cell similarity graphs in single cell omics \citep{Barkas2019-ck}, gene regulatory networks \citep{Bravo_Gonzalez-Blas2023-ra} and protein-protein interaction networks \citep{Cong2019-oc} are fundamental tools to rationalize the complexity of biological systems such as the human body. The Python and R interfaces of \texttt{igraph} are widely used in some of these areas and the improvements implemented in the last years are designed to create a solid basis for further expansion in these areas.

Despite its strengths, \texttt{igraph}'s design incurs some limitations as well. The current internal data structure is based on numbered arrays of vertices and edges, making it particularly efficient for static graph analysis but less performant for dynamic networks, i.e. networks in which vertices and edges are frequently added or removed. Although this is only noticeable on very large networks or for long simulations, future work is planned to provide a swappable core data structure that adapts to distinct use scenarios.

Another challenge of current \texttt{igraph} development is the entry barrier towards new code contributors. Although \texttt{igraph}'s codebase is vast and spans multiple languages, we are committed to creating a welcoming environment for new contributors. In particular, the joint operation with WINS is an exciting opportunity to include new contributors who identify with groups that have been historically underrepresented in \texttt{igraph}'s development team, such as women and non-binary people. As \texttt{igraph}'s coding standards continue to improve, it will be especially important to ``walk a few steps'' towards new community members in a context of mutual respect and positive encouragement. A new structure of formal governance, together with tighter interoperability with \texttt{networkx} and other packages, are key areas that will help increase accessibility for all types of users and developers.

\matmethods{

Scalability tests for Figure~\ref{fig:scaling} were performed on a tower workstation with 512 GB of RAM and using a single Intel Xeon Platinum 8160T CPU at 2.10GHz. Download statistics for PyPI were computed using PyPI stats (\url{https://pypistats.org/}). Dowload statistics for R were computed using cranlogs.app (\url{https://github.com/r-hub/cranlogs.app}). Number of edited lines, forks, and stars were computed using GitHub's online interface. Number of functions in Python were estimated using the inspect standard module. Number of functions in R were estimated using \texttt{lsf.str("package:igraph")}.

}

\showmatmethods{} 

\acknow{Development of \texttt{igraph} and its interfaces is supported by two Chan Zuckerberg Essential Open Source Software (EOSS) grants (N. EOSS2-0000000067 and EOSS4-0000000179) and a Chan Zuckerberg Diversity grant (N. EOSS-DI-0000000021). We would like to thank all \texttt{igraph} contributors for their valuable time and effort, and all members of the broader \texttt{igraph} community for their enthusiasm, engagement, and respectful attitude.}

\showacknow{} 

\bibsplit[3]

\bibliography{maintext}

\begin{thebibliography}{10}

\bibitem{erdos59a}
P Erd\"{o}s, A R\'{e}nyi, On random graphs i.
\newblock {\em\protect\JournalTitle{Publicationes Mathematicae Debrecen}}
  \textbf{6}, 290 (1959).

\bibitem{Barabasi1999-uz}
AL Barabasi, R Albert, Emergence of scaling in random networks.
\newblock {\em\protect\JournalTitle{Science}} \textbf{286}, 509--512 (1999).

\bibitem{Bankevich2012-um}
A Bankevich, et~al., {SPAdes}: A new genome assembly algorithm and its
  applications to {Single-Cell} sequencing.
\newblock {\em\protect\JournalTitle{J. Comput. Biol.}} \textbf{19}, 455--477
  (2012).

\bibitem{Tabula_Muris_Consortium2018-yg}
{Tabula Muris Consortium}, Single-cell transcriptomics of 20 mouse organs
  creates a tabula muris.
\newblock {\em\protect\JournalTitle{Nature}} \textbf{562}, 367--372 (2018).

\bibitem{Moreno1941-kk}
JL Moreno, Foundations of sociometry: An introduction.
\newblock {\em\protect\JournalTitle{Sociometry}} \textbf{4}, 15--35 (1941).

\bibitem{Burt2000-ne}
RS Burt, The network structure of social capital.
\newblock {\em\protect\JournalTitle{Research in Organizational Behavior}}
  \textbf{22}, 345--423 (2000).

\bibitem{Diani1992-bz}
M Diani, The concept of social movement.
\newblock {\em\protect\JournalTitle{Sociol. Rev.}} \textbf{40}, 1--25 (1992).

\bibitem{McFarland2014-iz}
DA McFarland, J Moody, D Diehl, JA Smith, RJ Thomas, Network ecology and
  adolescent social structure.
\newblock {\em\protect\JournalTitle{Am. Sociol. Rev.}} \textbf{79}, 1088--1121
  (2014).

\bibitem{networkx}
AA Hagberg, DA Schult, PJ Swart, Exploring network structure, dynamics, and
  function using networkx in {\em Proceedings of the 7th Python in Science
  Conference}, eds.{} G Varoquaux, T Vaught, J Millman.
\newblock (Pasadena, CA USA), pp. 11 -- 15 (2008).

\bibitem{peixoto_graph-tool_2014}
TP Peixoto, The graph-tool python library.
\newblock {\em\protect\JournalTitle{figshare}} (2014).

\bibitem{statnet2003}
MSH Pavel N.~Krivitsky, et~al., Statnet: Tools for the statistical modeling of
  network data (year?).

\bibitem{Shannon2003-qa}
P Shannon, et~al., Cytoscape: a software environment for integrated models of
  biomolecular interaction networks.
\newblock {\em\protect\JournalTitle{Genome Res.}} \textbf{13}, 2498--2504
  (2003).

\bibitem{d3js}
M Bostock, D3.js - data-driven documents (2012).

\bibitem{gephi}
M Bastian, S Heymann, M Jacomy, Gephi: An open source software for exploring
  and manipulating networks.
\newblock (2009).

\bibitem{Yen1970-pn}
JY Yen, An algorithm for finding shortest routes from all source nodes to a
  given destination in general networks.
\newblock {\em\protect\JournalTitle{Quart. Appl. Math.}} \textbf{27}, 526--530
  (1970).

\bibitem{Brandes2011-rw}
U Brandes, G Dorfm{\"u}ller, {PageRank} - what is really relevant in the
  {World-Wide} web? in {\em Algorithms Unplugged}, eds.{} B V{\"o}cking, et~al.
\newblock (Springer Berlin Heidelberg, Berlin, Heidelberg), pp. 89--96 (2011).

\bibitem{Gleich_2005-me}
D Gleich, L Zhukov, P Berkhin, Fast parallel {PageRank}: A linear system
  approach.
\newblock (2005).

\bibitem{horton}
JD Horton, A polynomial-time algorithm to find the shortest cycle basis of a
  graph.
\newblock {\em\protect\JournalTitle{SIAM Journal on Computing}} \textbf{16},
  358--366 (1987).

\bibitem{Wernicke2006-yz}
S Wernicke, F Rasche, {FANMOD}: a tool for fast network motif detection.
\newblock {\em\protect\JournalTitle{Bioinformatics}} \textbf{22}, 1152--1153
  (2006).

\bibitem{Ostergard2002-eg}
PRJ {\"O}sterg{\aa}rd, A fast algorithm for the maximum clique problem.
\newblock {\em\protect\JournalTitle{Discrete Appl. Math.}} \textbf{120},
  197--207 (2002).

\bibitem{Traag2019-xy}
VA Traag, L Waltman, NJ van Eck, From louvain to leiden: guaranteeing
  well-connected communities.
\newblock {\em\protect\JournalTitle{Sci. Rep.}} \textbf{9}, 5233 (2019).

\bibitem{McInnes2018-eu}
L McInnes, J Healy, J Melville, {UMAP}: Uniform manifold approximation and
  projection for dimension reduction.
\newblock {\em\protect\JournalTitle{arXiv}} (2018).

\bibitem{igraph1}
G Csárdi, T Nepusz, The igraph software package for complex network research.
\newblock {\em\protect\JournalTitle{InterJournal Complex Systems}}
  \textbf{1695} (2006).

\bibitem{Horvat2023-wh}
S Horv{\'a}t, et~al., {IGraph/M}: graph theory and network analysis for
  mathematica.
\newblock {\em\protect\JournalTitle{J. Open Source Softw.}} \textbf{8}, 4899
  (2023).

\bibitem{Blondel2008-ht}
VD Blondel, JL Guillaume, R Lambiotte, E Lefebvre, Fast unfolding of
  communities in large networks.
\newblock {\em\protect\JournalTitle{J. Stat. Mech.}} \textbf{2008}, P10008
  (2008).

\bibitem{fluidcommunities}
F Par{\'e}s, et~al., Fluid communities: A competitive, scalable and diverse
  community detection algorithm in {\em Complex Networks {\&} Their
  Applications VI}, eds.{} C Cherifi, H Cherifi, M Karsai, M Musolesi.
\newblock (Springer International Publishing, Cham), pp. 229--240 (2018).

\bibitem{Cox2008-lr}
MAA Cox, TF Cox, Multidimensional scaling in {\em Handbook of Data
  Visualization}, eds.{} CH Chen, W H{\"a}rdle, A Unwin.
\newblock (Springer Berlin Heidelberg, Berlin, Heidelberg), pp. 315--347
  (2008).

\bibitem{Adai2004-ba}
AT Adai, SV Date, S Wieland, EM Marcotte, {LGL}: creating a map of protein
  function with an algorithm for visualizing very large biological networks.
\newblock {\em\protect\JournalTitle{J. Mol. Biol.}} \textbf{340}, 179--190
  (2004).

\bibitem{Davidson1996-vz}
R Davidson, D Harel, Drawing graphs nicely using simulated annealing.
\newblock {\em\protect\JournalTitle{ACM Trans. Graph.}} \textbf{15}, 301--331
  (1996).

\bibitem{Frick1995-eg}
A Frick, A Ludwig, H Mehldau, A fast adaptive layout algorithm for undirected
  graphs (extended abstract and system demonstration) in {\em Graph Drawing}.
\newblock (Springer Berlin Heidelberg), pp. 388--403 (1995).

\bibitem{Granger2021-rz}
BE Granger, F P{\'e}rez, Jupyter: Thinking and storytelling with code and data.
\newblock {\em\protect\JournalTitle{Comput. Sci. Eng.}} \textbf{23}, 7--14
  (2021).

\bibitem{Hunter:2007}
JD Hunter, Matplotlib: A 2d graphics environment.
\newblock {\em\protect\JournalTitle{Computing in Science \& Engineering}}
  \textbf{9}, 90--95 (2007).

\bibitem{plotly}
PTI Plotly, Collaborative data science (2015).

\bibitem{mckinney-proc-scipy-2010}
{W}es {M}c{K}inney, {D}ata {S}tructures for {S}tatistical {C}omputing in
  {P}ython in {\em {P}roceedings of the 9th {P}ython in {S}cience
  {C}onference}, eds.{} {S}t\'efan van~der {W}alt, {J}arrod {M}illman.
\newblock pp. 56 -- 61 (2010).

\bibitem{crandep}
C Lee, Modelling the number of reverse dependencies (2023).

\bibitem{pkgdown}
H Wickham, J Hesselberth, M Salmon, {\em pkgdown: Make Static HTML
  Documentation for a Package}, (2022) https://pkgdown.r-lib.org,
  https://github.com/r-lib/pkgdown.

\bibitem{oscar-najera}
Óscar Nájera, et~al., sphinx-gallery/sphinx-gallery: v0.12.2 (2023).

\bibitem{Kwak2010-zh}
H Kwak, C Lee, H Park, S Moon, What is twitter, a social network or a news
  media? in {\em Proceedings of the 19th international conference on World wide
  web}, WWW '10.
\newblock (Association for Computing Machinery, New York, NY, USA), pp.
  591--600 (2010).

\bibitem{sinha2015an}
A Sinha, et~al., An overview of microsoft academic service (mas) and
  applications in {\em International World Wide Web Conferences}.
\newblock (Microsoft), (2015).

\bibitem{Barkas2019-ck}
N Barkas, et~al., Joint analysis of heterogeneous single-cell {RNA-seq} dataset
  collections.
\newblock {\em\protect\JournalTitle{Nat. Methods}} \textbf{16}, 695--698
  (2019).

\bibitem{Bravo_Gonzalez-Blas2023-ra}
C Bravo Gonz{\'a}lez-Blas, et~al., {SCENIC+}: single-cell multiomic inference
  of enhancers and gene regulatory networks.
\newblock {\em\protect\JournalTitle{Nat. Methods}} \textbf{20}, 1355–--1367
  (2023).

\bibitem{Cong2019-oc}
Q Cong, I Anishchenko, S Ovchinnikov, D Baker, Protein interaction networks
  revealed by proteome coevolution.
\newblock {\em\protect\JournalTitle{Science}} \textbf{365}, 185--189 (2019).

\end{thebibliography}

\end{document}